\documentclass[runningheads]{llncs}
\usepackage[T1]{fontenc}
\usepackage{graphicx}
\usepackage{float} 
\usepackage{xcolor}
\begin{document}
\title{PrivSTRUCT: Untangling Data Purpose Compliance of Privacy Policies in Google Play Store}
\titlerunning{PrivSTRUCT}
%
%
\author{Bhanuka Silva\inst{1}\orcidID{0009-0000-6558-6514} \and
Anirban Mahanti\inst{1}\orcidID{0009-0003-9030-9916} \and
Aruna Seneviratne\inst{2}\orcidID{0000-0001-6894-7987} \and
Suranga Senevirante\inst{1}\orcidID{0000-0002-5485-5595}}
\authorrunning{Silva B. et al.}
%
\institute{The University of Sydney, Australia\\ 
\email{\{bpin9254,anirban.mahanti,suranga.seneviratne\}@sydney.edu.au} \and
University of New South Wales, Australia\\
\email{a.seneviratne@unsw.edu.au}}
\maketitle              
\begin{abstract}
Existing research typically treats privacy policies as flat, uniform text, extracting information without regard for the document's logical hierarchy. Disregard for structural cues of section headings designed to guide the reader, often leads automated methods to entangle distinct data practices, particularly when linking sensitive data items to their specific purposes. To address this, we introduce \textbf{PrivSTRUCT}, a novel and systematic encoder and decoder combined framework that to untangle complex privacy disclosures. Benchmarking against the state-of-the-art tool PoliGrapher reveals that PrivSTRUCT robustly extracts more than \textbf{x2} the number of data item and purpose excerpts while retaining developer-defined structural cues. By applying PrivSTRUCT to a large-scale dataset of 3,756 Android apps, we uncover a critical transparency gap: the probability of developers overstating a data purpose is \textbf{20.4\%} higher for first-party collection and \textbf{9.7\%} higher for third-party sharing when they rely on globally defined purposes rather than specific, locally scoped disclosures. Alarmingly, we find that sensitive third-party data flows such as sharing financial data for analytics are frequently diluted and entangled into generic or unrelated categories, highlighting a persistent failure in the current purpose disclosure landscape.

\end{abstract}
\section{Introduction}
\label{5sec:introduction}

Privacy policies are to inform users about data collection and usage practices, yet they are notoriously complex and lengthy, often leading individuals to skip reading them altogether \cite{cost_of_reading,www21_PP_overtime,why_ignore_pp}. Regulations such as the EU General Data Protection Regulation (GDPR), the California Consumer Privacy Act (CCPA), and the Australian Privacy Principles (APPs) have sought to improve transparency. However, these efforts have paradoxically increased policy intricacy, making them even less accessible \cite{take_some_cookies,policy_landscape}. In the mobile app ecosystem, where users increasingly rely on applications for daily tasks, this issue is amplified. App marketplaces like the Google Play Store require developers to provide summarised data practices through Android data safety labels \cite{Google_privacy_policy}, which outline collected or shared data items and their purposes in a more straightforward manner. However, these self-declared labels must align with the legally binding full privacy policy document in order to avoid misleading users and ensure compliance \cite{GDPRthirdParty}. Yet, as we demonstrate in this work, such alignments are frequently undermined by structural ambiguities in the policies themselves.

\begin{figure}[t]
    \centering
    \includegraphics[width=0.99\textwidth]{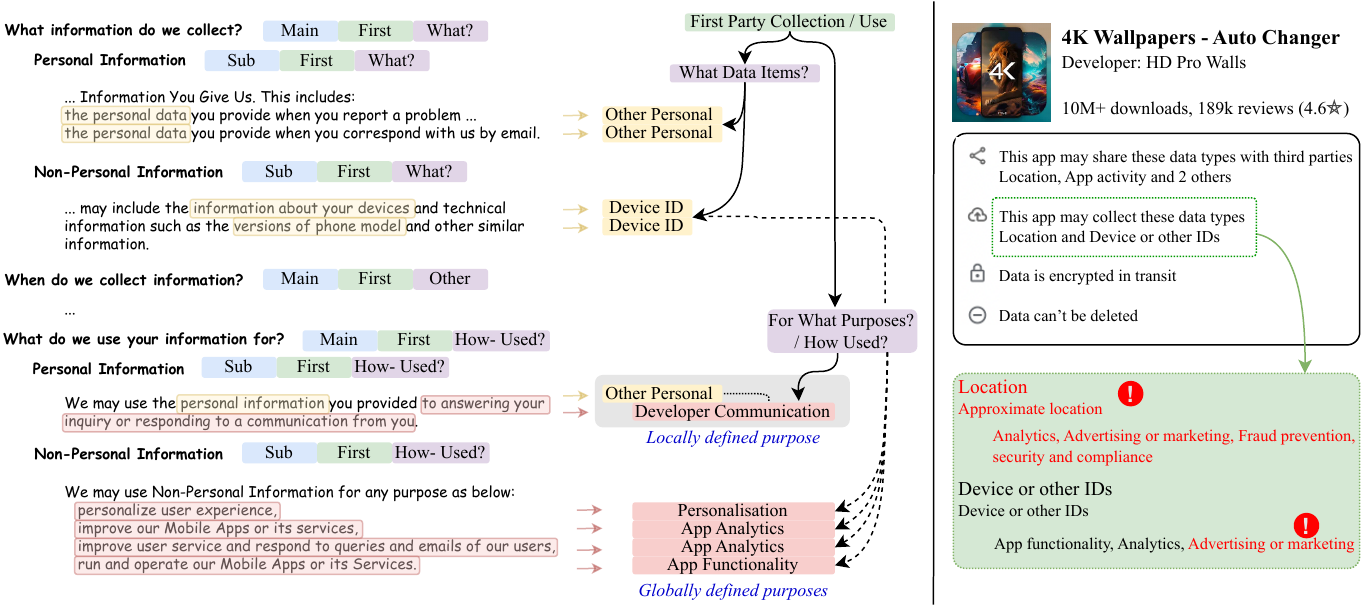}
    \caption[Walkthrough example]{Left: Privacy policy portion of `HD Pro Walls' developer related to first party collection details. Right: The data safety declaration of the developer's app called `4K Wallpapers - Auto Changer', with nearly 10M+ downloads, yet contradicting with the policy itself based on data practices. Extracted on December, 2025}
    \label{5fig:walkthrough_example}
\end{figure} 

Empirical observations reveal that many policies separate descriptions of data items (e.g., “name” or “email address”) from their purposes (e.g., “app analytics” or “personalisation”), often across distinct sections. This isolation can prompt users to make blanket consents, assuming broad applicability without clear linkages, potentially leading to unfair data processing. 
Figure \ref{5fig:walkthrough_example} depicts ``locally-defined'' purposes where specific data like \emph{personal information} is explicitly linked to \emph{developer communication} and ``globally-defined'' purposes where ambiguous text forces readers to guess and map  disconnected data items like \emph{device IDs} to the purposes such as \emph{app analytics, functionality and personalisation} defined in later sections of the privacy policy. This figure also depicts how assigning meaningful labels to the main and sub section headings can structurally link disjoint elements. In this example, data items related to \emph{first-party collection / use} are globally mapped with homogeneous data purpose declarations based on these labels. In the right is the data safety declaration of the most popular app by this developer with nearly 10M+ downloads, which alarmingly indicates that they collect approximate location from end-users for \emph{analytics, advertising, fraud prevention security and compliance} but this is not defined at all in the privacy policy, indicating a major contradiction when regulatory frameworks require consistency across both forms of disclosure. Additionally, use of \emph{device IDs} for \emph{advertising or marketing} is also not disclosed in policy as well.

Existing natural language processing (NLP) approaches often exacerbate this issue by either analysing policies as flat sequences of text, or disregarding the hierarchical cues (such as section headings) that developers use to scope these claims, or by not considering the data purposes altogether. 
Consequently, these methods risk extracting undistinguished relationships that misrepresent the true granularity of the developer's disclosure.
To this end, we propose PrivSTRUCT (\underline{Priv}acy policy \underline{S}tructural \underline{T}agging for \underline{R}obust \underline{U}nderstanding and \underline{C}ompliance \underline{T}racing), a novel framework that resolves policy ambiguity by systematically leveraging the developer-intended structural hierarchy. PrivSTRUCT utilises a hybrid architecture combining decoder-based high-level structural extraction with encoder-based classification for granular semantic analysis. This allows us to not only map the logical flow of a policy but to specifically categorise section headings by intent, distinguishing between segments that define what data is collected, why it is processed, why it is shared, etc. By reconstructing these dependency links, PrivSTRUCT can accurately differentiate between locally defined purposes (tied to specific data items) and globally defined mandates, enabling a far more precise audit of Play Store Data Safety Labels than previously possible. More specifically, our contributions include:

\begin{itemize} 
\item PrivSTRUCT, a novel systematic framework that combines the natural language understanding (NLU) capabilities of decoder-based LLMs with the classification-efficiency of encoder-based models to analyse privacy policies. It leverages structural cues to untangle the relationships between data items and purposes. We introduce a Direct Preference Optimisation (DPO) pipeline to fine-tune smaller, open-source models (Llama-3.1) for structural extraction, achieving performance comparable to proprietary state-of-the-art models while significantly reducing computational overhead.

\item We benchmark PrivSTRUCT against \emph{PoliGraph}, a state-of-the-art NLP tool for privacy policy analysis. Our evaluation on a diverse test set reveals that in average PoliGraph only identifies 52.1\% and 89.1\% less number of unique data items and data purposes compared to our method.

\item We conduct a large-scale empirical analysis of 3,756 privacy policies and their corresponding Google Play Data Safety labels \footnote{Dataset is available via \url{https://github.com/NSS-USYD/PrivCORPUS/}}. We characterise the landscape of ``Purpose Compliance'' and ``Purpose Dilution'', revealing that the probability of developers overstating purposes is 20.4\% higher for first-party collection and 9.7\% higher for third-party sharing as they increasingly rely on globally defined purposes. Alarmingly, we find that critical third-party data flows such as sharing financial data for analytics are frequently diluted into generic or unrelated categories within the policy text, undermining the transparency goals of modern app marketplaces. 
\end{itemize}


\section{Related Work}
\label{5sec:literature}

Prior efforts to demystify privacy policies relied on traditional NLP for summarisation and decision support \cite{gopinath2020automatic,sathyendra2016automatic,sathyendra2017identifying,Polisis_usenix,zimmeck2014privee,nokhbeh2020privacycheck}. These methods were eventually superseded by encoder-based transformers and related architectures \cite{adhikari2023evolution,srinath2021privacy}. More recently, the landscape has shifted toward generative approaches, which are increasingly demonstrating remarkable efficacy in zero- and few-shot policy understanding \cite{tang2023policygpt,rodriguez2024largelanguagemodelsnew,silva2024entailment,chanenson2025automating}.

Extracting structured information from policy text via semantic analysis is crucial for tasks such as consistency checking. PolicyLint \cite{policylint} introduced linguistic analysis to extract (data type, entity) tuples from individual sentences, enabling the detection of internal contradictions. Building on this, PurPliance \cite{bui2021consistency} extended the tuple structure to identify processing purpose clauses, allowing for comparisons against actual app behaviour. Similarly, PolicyChecker \cite{andow2020actions} leveraged semantic analysis roles to audit the consistency between policy disclosures and app code. However, these approaches rely on isolated sentence-level analysis; they cannot resolve dependencies across multiple sections of a policy, often resulting in extracted tuples that are disconnected or ambiguous. Alternative frameworks like PolicyComp \cite{zhou2023policycomp} attempt to bypass these extraction challenges by benchmarking policies against those of similar `counterpart' apps, but this merely identifies statistical outliers and does not resolve the fundamental inability of current models to handle complex, cross-section dependencies

 Recent work PoliGraph \cite{cui2023poligraph} moves beyond linear text analysis by modelling privacy policies as knowledge graphs. In this architecture, data types and entities function as nodes, connected by two primary edge relations: SUBSUME, which maps generic terms to specific definitions, and COLLECT, which links entities to the data they acquire. PoliGraph identifies 40\% more collection statements than prior state-of-the-art while achieving 97\% precision, however, `purposes' of such collections are treated merely as a secondary attribute attached to the COLLECT edge. 
 This dependency presents a significant limitation: purposes in complex policy text are frequently ambiguous or syntactically detached from explicit collection verbs. As a result, purposes are not always well-defined solely by the existence of a link between an entity and a data type. This structural rigidity often leads to missed or misattributed purposes, underscoring the need for more robust identification methods. PrivPRISM \cite{silva2026privprism} is a LLM decoder and encoder combined framework specifically designed to extract data items and purposes for compliance auditing. Despite its efficiency for robust data extraction, 
it does not explicitly infer non-trivial relationships between data items and purposes. PrivPRISM introduces section heading extraction as an integral part of the framework but do not deep dive into intentions of the headings. To address these limitations, the PrivSTRUCT framework introduced in this paper leverages previously ignored developer-disclosed intentions within section headings. By modelling this high-level policy flow, PrivSTRUCT establishes robust links between disconnected subsections, enabling the precise mapping of otherwise ambiguous, globally-defined purposes.


\section{PrivSTRUCT Framework}
\label{5sec:methodology}



\begin{figure}[t]
    \centering
    \includegraphics[width=0.99\textwidth]{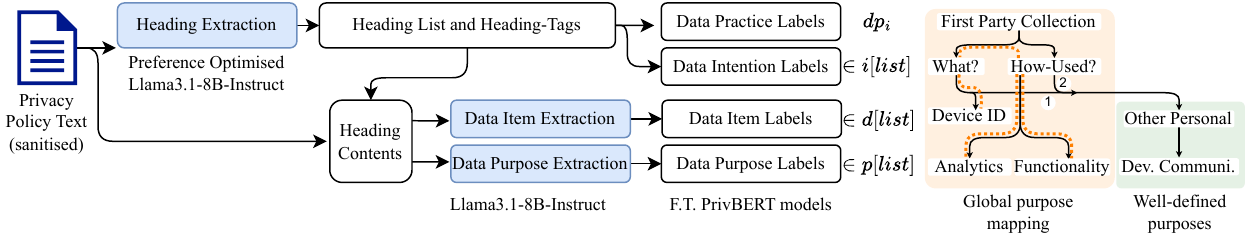}
    \caption{PrivSTRUCT framework}
    \label{5fig:framework}
\end{figure}

This section first details the privacy policy dataset, followed by a breakdown of each module within the PrivSTRUCT framework (the complete pipeline is illustrated in Figure \ref{5fig:framework}). Finally, we introduce the metrics used to characterise data purpose compliance and purpose dilution.



\subsection{Dataset}
\label{5subsec:dataset}

We analyse a corpus of 3,756 unique privacy policies associated with 6,540 of the most downloaded apps on the Google Play Store, crawled during the first half of 2024. By restricting our analysis to policies with a sanitised text file size < 50KB, we capture 97\% of the initially observed top policies while excluding statistical outliers (the corpus average is 18.3KB). Focusing on top-ranked applications aligns with the findings of Khandelwal et al. \cite{khandelwal2024unpacking}, who note that developers of high-visibility apps typically provide more robust declarations due to heightened public scrutiny and available resources. Notably, this dataset reveals a widespread pattern of policy reuse; major developers frequently govern multiple high-traffic apps under a single, unified text. To ensure accurate compliance attribution, we map each unique privacy policy strictly to the developer's most downloaded application.


Categorically, the dataset is dominated by \emph{tools} (20.2\%), \emph{photography} (8.5\%), and \emph{entertainment} (8.4\%). The high proportion of utility apps makes this corpus particularly relevant for our study, as these apps frequently necessitate broad Android system permissions that are functionally justified yet vaguely articulated in generic policy texts. Personal data sensitive categories like \emph{finance} (3.3\%) and \emph{medical} (0.6\%) represent a much smaller fraction, reflecting their specialised, rather than general-purpose, user base.

To systematically develop the PrivSTRUCT framework, we extracted a representative subset of 750 policies ($DS_{dev}$). To mitigate potential bias toward the length, we applied a stratified sampling technique, guaranteeing a uniform distribution by selecting exactly 150 policies from each 10KB length increment. This subset was subsequently divided into 675 policies for training ($DS_{train}$) and 75 for testing ($DS_{test}$). Following the finalisation of the PrivSTRUCT model, we executed our at-scale empirical compliance analysis across the entire 3,756 policy corpus.

\subsection{Heading extraction}
\label{subsec:heading_extraction}

\begin{figure}[t]
    \centering
    \includegraphics[width=0.8\textwidth]{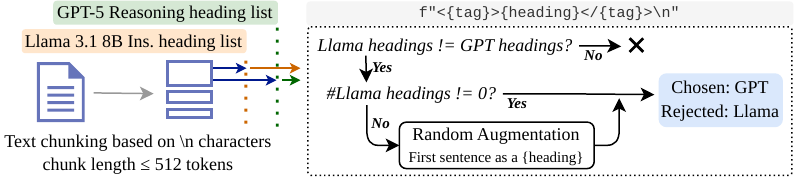}
    \caption{Synthesising of the DPO dataset}
    \label{fig:dpo_dataset_create}
\end{figure}

Modern HTML based privacy policies frequently rely on complex CSS and JavaScript rather than a semantic Document Object Model (DOM) tree, therefore, developers rarely use standard HTML heading tags (e.g., \texttt{<h1>} through \texttt{<h6>}) \cite{opt_outs_from_policy,Polisis_usenix,silva2026privprism}. Using standard HTML parsers then inevitably reduce the document to a flat text output, stripping away implicit hierarchical cues. To recover this lost structure, we feed the flattened text file into a LLM tasked with identifying and reconstructing the heading hierarchy.
To determine the most efficient approach for at-scale extraction, we compare the proprietary GPT-5 reasoning (400k context window) via API against the open-weight Llama-3.1-8B-Instruct (128k context window), evaluating the latter for its potential cost and inference-time advantages in local deployments. Across $DS_{dev}$, we prompt both models to generate a structural flow from the flat text, explicitly annotating each identified heading with either a \texttt{<main>} or \texttt{<sub>} tag to indicate its hierarchical level (c.f. Figure \ref{fig:dpo_dataset_create}).



While both models performed similarly on the majority of policies, empirical evaluations revealed three consistent failure modes for the Llama model: (1) defaulting to uniform heading levels (all main or all sub), (2) misclassifying entire paragraph sentences as headings, and (3) dropping existing headings entirely. To address this without the extensive overhead of traditional supervised fine-tuning, we employed Direct Preference Optimisation (DPO) \cite{rafailov2023direct} to align the Llama model with GPT-5's extraction quality.
As depicted in Figure \ref{fig:dpo_dataset_create}, we curated a DPO dataset from $DS_{train}$ by chunking them into 512-token segments and isolating the instances where Llama's baseline extraction mismatched GPT-5's output (71.4\% of 8,418). To explicitly penalise Llama's tendency to extract full sentences (failure mode 2), among the missed-heading instances (28.9\%) we augmented 50\% of them by supplying the segment's first non-heading sentence as a target rejected output. 
We use $DS_{test}$ to evaluate the DPO training, specifically tuning the deviation penalty hyper-parameter, $\beta$, to optimise heading extraction quality. To quantify this quality, we measure the "fitment" of headings across the policy by calculating the median and inter-quartile range (IQR) of the text length between identified headings. This distribution effectively flags structural errors; for instance, a lower quartile approaching zero indicates the model is likely misidentifying consecutive list items as sub-headings. We also track the average number of headings identified per policy to ensure comprehensive extraction, with detailed results discussed in Section \ref{subsec:benchmark_dpo}.

\subsection{Decoder based data item and purpose extraction}
Literature highlights that embedding based models are not effective in semantic text excerpt search due to the vector space noise \cite{gao2023retrieval}. Comparatively it is straightforward for decoder models as a natural language understanding (NLU) task \cite{grattafiori2024llama3herdmodels,silva2026privprism} to extract which segments represent data items and data purposes for a given text chunk with 512 tokens at most. Given the nature of this task, we directly use Llama3.1 8B instruct and represents outputs in a JSON format;
$$[\{`data':`data~excerpt', `purpose':`purpose~excerpt'\},...]$$ and we emphasise that we allow data or purpose keys to contain empty strings if necessary. If we observe any empty purpose for a given data, we later on try to map them globally with a suitable purpose discussed elsewhere, which is a key novelty in this work.

To link \textbf{globally-defined} purposes, we create a link between isolated purposes (i.e. no locally defined data item) listed under a \emph{data intention = Why? or How Used?} and isolated data items (i.e. no locally defined data purpose) listed under a \emph{data intention = What?}. An example can be observed in orange colour dashed lines in Figure \ref{5fig:framework}. If we observe isolated data items or purposes not belonging to above mentioned data intentions, we still link them but as \textbf{weak global (floating) relationships.} Note that the above mentioned two linkages are pre-conditioned on \emph{data practice} category and we do not match first party collection data items with third party sharing data purposes (or vice versa).

\subsection{Encoder based classifiers}

Following the information extraction phase, we employ four fine-tuned PrivBERT models to classify the extracted text into their respective categories. PrivBERT is a privacy policy pre-trained RoBERTa based embedding model (c.f. Sub Section \ref{subsec:benchmarking_structural_cues}). The fine-tuning of these modules for the downstream task of label prediction was based on synthetic labels generated via GPT5 prompting for information extracted from the 750 privacy policies. This is similar to LLM bootstrapped training of encoders discussed in \cite{silva2026privprism}. We demonstrate the fine-tuning performance in Table \ref{tab:encoder_results}. 

The classifiers for Data Item and Data Purpose operate on a straightforward paradigm: they map each input text excerpt to a single class label using batched inferencing. While the Data Practice and Data Intent classifiers could follow this same approach, we hypothesise that encoder-based models benefit significantly from additional structural context. To test this, we evaluate an input schema that includes the section heading list within the prompt itself. Specifically, we format the input as ``\texttt{[CLS] target\_heading [SEP] context\_headings [SEP]}''. Due to PrivBERT’s 512-token input limit, we include only a subset of adjacent headings (a `neighbourhood') rather than the full list. We evaluate the impact of this context-aware approach in Sub Section \ref{subsec:benchmark_encoder}. A complete list of classification labels given below.

\begin{itemize}
\scriptsize
\item \textbf{Data Practices:} First Party Collection/Use ($c_0$), Third Party Sharing ($c_1$), User Choice/Control ($c_2$), User Access, Edit and Deletion ($c_3$), Introductory/Generic ($c_4$), Policy Change ($c_5$), Data Security ($c_6$), International and Specific Audience ($c_7$), Practice Not Covered ($c_8$), Data Retention ($c_9$), Privacy Contact Information ($c_{10}$), and Do Not Track ($c_{11}$).
\item \textbf{Data Intentions:} What ($i_0$), Why ($i_1$), How-Collected ($i_2$), How-Used ($i_3$), When ($i_4$), Other ($i_5$).
\item \textbf{Data Items ($D$):} Name ($d_0$), Email ($d_1$), User account ($d_2$), Address ($d_3$), Phone ($d_4$), Race/Ethnicity ($d_5$), Political/Religious ($d_6$), Gender ($d_7$), Financial ($d_8$), Location ($d_9$), Search and Browsing history ($d_{10}$), SMS/ Messages/ Call log ($d_{11}$), Photos/Videos ($d_{12}$), Audio/Music ($d_{13}$), Health/Fitness ($d_{14}$), Contacts ($d_{15}$), Calendar ($d_{16}$), App performance/ App Activity ($d_{17}$), Device identifier ($d_{18}$), Files/Documents ($d_{19}$), Other Personal ($d_{20}$), Generic information ($d_{21}$) and Negatives ($d_{22}$).
\item \textbf{Data Purposes ($P$):} App Analytics ($p_0$), Developer communication ($p_1$), Fraud prevention / security and compliance ($p_2$), Advertising or marketing ($p_3$), Personalisation ($p_4$), Account management ($p_5$), App functionality ($p_6$), and Other ($p_7$).   
\end{itemize}

\subsection{Metrics for Data Purpose Compliance}
\label{subsec:metrics-compliance}

In this subsection, \emph{compliance} indicates whether a given data item and its purpose is disclosed in the data safety (DS) declarations and also correctly reflected and described in the corresponding privacy policy (PP) text. More formally,  for a given $\{PP, DS\}$ pair, we could observe the purpose compliance by comparing the presence of each individual purpose for and when a given data item is agreed as collected or shared among the pair. 
To establish a generalised landscape of data purpose compliance (for all pairs of PP and DS instances), we define the probability of observing a given purpose ($p[i]$) being \emph{well-disclosed} for a given data item ($d[j]$) as $P_{WD}(i,j)$ if and only if \textbf{locally-defined}. 

\begin{footnotesize}
$$P_{WD}(i,j) = P(p[i] \in (PP ~\cap~ DS) ~|~ d[j] \in (PP ~\cap~ DS)) $$   
\end{footnotesize}

$P_{OS}(i,j)$ represents the probability when a purpose $p[i]$ that is \textbf{locally} defined in the privacy policy is not observed in data safety, and we consider this as the probability of \emph{purpose over-statement}. Similarly $P_{US}(i,j)$ which is the probability of \emph{purpose under-statement} defines that a purpose is mentioned in the data safety but not in the privacy policy. (Note that the condition of purpose non-existence or $P_{NE}(i,j)$ is not explicitly mentioned due to less significance but can be obtained by $1 - P_{WD}(i,j) - P_{OS}(i,j) - P_{US}(i,j)$).

\begin{footnotesize}
    $$P_{OS}(i,j) = P(p[i] \in (PP \setminus DS) ~|~ d[j] \in (PP ~\cap~ DS)) $$ 
    $$P_{US}(i,j) = P(p[i] \in (DS \setminus PP) ~|~ d[j] \in (PP ~\cap~ DS)) $$  
\end{footnotesize}
Note: Probabilities are computed per data item $d[j]$ of a given data practice classification type. In the main context of the chapter, we are mostly interested about data practices of PPs related with first party collection ($c_0$) and third party sharing ($c_1$) which can be directly compared and contrasted against DS declarations regarding compliance. 

As we have discussed before, developers are inclined to not locally define data item and data purpose relationships, and rather they tend to disclose via \textbf{global definitions}. I.e. data items may be in bulk declared in one section related to collection practices with relevant purposes declared elsewhere and vice versa for sharing practices. As PrivSTRUCT is capable of extracting such relationships, it is important to evaluate the compliance when taken these into consideration as well. The \textbf{delta} contribution we receive when these \textbf{globally-defined} data-item and purpose pairs are considered to the original $P_X$ is denoted as $\Delta$. For example, based on locally and globally defined data items and purposes, probability of well-disclosed purposes being observed will be $P_{WD}+\Delta P_{WD}$. 

For us to globally match such purposes, it is expected that the developers disclose data items inside a heading related to ``WHAT" data intention (e.g. What information do we collect from you?) and disclose the data purposes inside a heading related to ``WHY" data intention (e.g. Why do we collect information from you?). However, developers could still disclose \textbf{floating} data items and data purposes in other sections where the true intent is not clear. In this scenarios, when the data practice category is strictly clear (i.e. collection or sharing), we \textbf{weakly global match} such data items and data purposes and the additional contribution we receive from this mapping to the compliance is indicated by $\Delta'$. 


\subsection{Metrics for Data Purpose Dilution}
\label{subsec:metrics-dilution}

Data Purpose Dilution refers to a phenomenon where the purpose for collecting or sharing a specific data item, as declared in the DS declaration, does not directly align with the purpose stated in the PP. Instead, the purpose in one document (e.g., DS) is “diluted” or spread across a diverse set of other purposes in the other document (e.g., PP) for the same data item, or vice versa. This misalignment creates ambiguity and reduces clarity for end-users about why their data is being processed.

To calculate the data purpose dilution matrix ($DM$) we deploy the logic given by the following equation where we identify the disagreements between the purposes of DS and PP for a given data item $j$ and dilute them across other purposes with disagreements. We use a 8x7 matrix as the `other' purpose is non existent for DS declaration for which it always follow $p[7] \in DS = 0$, 

\begin{footnotesize}
    $$DM(x,y) = (p[x] \in PP \setminus DS) \land (p[y]\in DS \setminus PP)$$
\end{footnotesize}

We normalise each dilution matrix for a given data type for the summation of all cells to be one for a given DS and PP pair. For observing the general landscape of data dilution, we add all dilution matrices (per data item) and normalise again. (Note: DM is different from the confusion matrix in traditional multi-class classification problems, as there are no set ground truth and prediction values for the problem we are discussing.) This metric helps us to identify which data purposes for a given data item are consistently not-disclosed by the developers while disclosing one or more unrelated data purposes. 

\section{Benchmarking PrivSTRUCT}
\label{5sec:benchmarking}

To validate the efficacy of the PrivSTRUCT framework, we conduct a series of benchmarking analysis starting with confirming the availability of structural cues. Next the capabilities of DPO-trained models against state-of-the-art LLM baselines and next, the classification performance of encoder-based modules under varying context settings. Finally we compare PrivSTRUCT with PoliGrapher, a state-of-the-art knowledge graph based NLP tool for extracting privacy policy information.

\subsection{Confirming the availability of structural cues via headings}
\label{subsec:benchmarking_structural_cues}

Meaningful label assignment via privacy policy paragraph classification is widely addressed in literature, particularly, OPP-115 foundational dataset introduced by Wilson et al. \cite{opp115} in which legal experts annotated 23k segments belonging to 115 online privacy policies has become popular and widely adapted as in \cite{sathyendra2016automatic,story2019natural,Polisis_usenix,srinath2021privacy,silva2024entailment}. PrivBERT, a RoBERTa based encoder model that is pre-trained on millions of privacy policies and fine-tuned on OPP-115 excels in this classification task \cite{srinath2021privacy,silva2024entailment}. Selection of the labels $c_0$ to $c_{11}$ in this work overlaps with this line of prior work.

However, we rarely see a privacy policy written as a continuous block of text that require individual paragraphs to be classified, rather, developers use section headings to direct end-users to specific parts of the policy. In this benchmarking experiment, we evaluate \emph{whether (a) section heading based (no paragraph context given) labels align with (b) paragraph based (no section headings given) embedding model generated labels.} As there are no existing embedding models trained for (a), we employ the GPT-5 reasoning model over $DS_{dev}$ to extract all the headings from privacy policy text files and to assign each heading with two levels of labels: data practice tags (i.e. $c_0$ to $c_{11}$) and data intention tags (i.e. $i_0$ to $i_5$). For (b) we directly use the most confident label provided by the fine-tuned PrivBERT classifier.

\begin{figure}[t]
    \centering
    \includegraphics[width=0.99\textwidth]{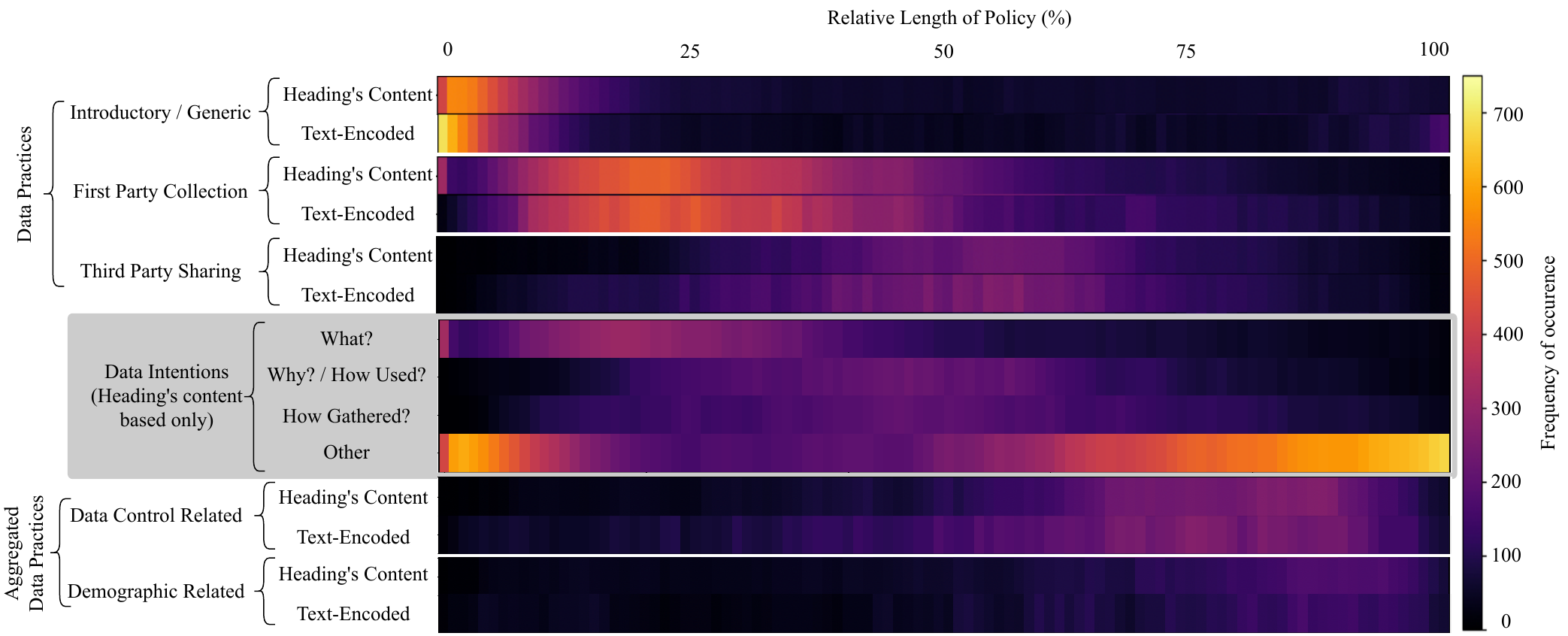}
    \caption[Text encoder vs heading-content based classification]{Analysing 750 privacy policies based on semantic (text encoder) classification versus structural (heading-content basis) labels}
    \label{5fig:background}
\end{figure} 

In the results depicted in Figure \ref{5fig:background}, x-axis represents the relative position within the privacy policy, normalised on a scale of 0.0 (start of the document) to 1.0 (end of the document). Comparing the distribution of labels derived from text-encoding versus those derived purely from headings, a near-identical distributions across both methods are observed. This result highlights a common distributional pattern across privacy policies. The initial 10\% of the policy length is typically introductory or generic. Next, nearly 40\% of the document length is dedicated to \emph{First Party Collection}. In this segment, we observed a relative agreement of 98.65\% between the heading-based and text-embedding based methods. The subsequent 25\% of the document predominantly details \emph{Third Party Sharing}, with a method agreement of 91.65\%. \emph{Data Control} practices (an aggregation of user choice, access / edit / deletion, retention, and security) and \emph{Demographic} explanations (e.g., target regions like EU or US) are predominantly located in the final quartile.
Proven the hypothesis on alignment is valid, finally, the GPT-5 based \emph{data-intention-tags} reveal that sections specific to ``why'' data is collected or ``how'' it is used are structurally distinct from sections describing ``what'' data is collected. This confirms that developers implicitly structure policies with separated intents, a cue that can be leveraged for more granular analysis, \emph{an important feature not utilised before in literature.}

\subsection{DPO for heading extraction}
\label{subsec:benchmark_dpo}

Based on the experiments outlined in Section \ref{subsec:heading_extraction}, a hyperparameter selection experiment yielded $\beta=0.4$ as the optimal value where the DPO trained Llama3.1 model provided highest quality of outputs related to heading structure extraction. We refer the readers to Appendix \ref{appendix:DPO} for detailed results on this hyperparameter selection and further explanations. A manual study over $DS_{test}$ further confirmed preference for DPO-Llama3.1 model improved to 42.7\%, a performance that is highly competitive with respect to the proprietary GPT-5 (50.7\%) and baseline Llama3.1 (6.6\%). It confirms that we can reliably substitute a locally hosted decoder model for heading extraction with no API costs and much faster inference times (c.f. Appendix Figure \ref{fig:dpo_results} - d). This preference optimised model was then used for generating empirical results for full dataset.

\subsection{Encoder based classifiers}
\label{subsec:benchmark_encoder}

\begin{table}[t]
\centering
\caption{Macro average classification results of the four classifiers}
\begin{footnotesize}
\begin{tabular}{llcccccc}
\hline
Classifier & Feed &\#Labels & \#Train & \#Test & Pr & Re & F1 \\
\hline
1. Data Practice &  single     & 12& 21k & 2795    & 0.86 & 0.86 & 0.86 \\
                 &  multiple   & 12& 21k & 2795    & 0.96 & 0.96 & 0.96 \\

2. Data Intention & single      & 6  & 21k & 2795     & 0.85 & 0.85 & 0.85 \\
                  & multiple    & 6  & 21k & 2795    & 0.87 & 0.87 & 0.87 \\

3. Data Items &  N/A      &23&74k & 8271    & 0.87 & 0.83 & 0.85 \\
4. Data Purpose &  N/A    & 8&51k & 6320       & 0.93 & 0.92 & 0.93 \\

\hline
\\   
\end{tabular}
\end{footnotesize}
\label{tab:encoder_results}
\end{table}

Text classification using encoder models is well-established and proven efficient for privacy policy analysis \cite{srinath2021privacy,silva2024entailment}. Table \ref{tab:encoder_results} presents the macro-average scores for the four classifiers utilised in PrivSTRUCT.
The column \#Train indicates how many samples we used to fine tune each model and we followed LLM bootstrapped training, where the training labels are generated via GPT5 models. This stage allows us to use efficient encoder models ($\sim$500M parameters in total) for the label generation task without relying on locally hosted much larger Llama3.1 models or API based generative models, still providing comparable performance.

We specifically highlight the ablation studies for the data practice and data intention classifiers (rows 1 and 2 in Table \ref{tab:encoder_results}). Here, `feed' denotes whether the model received only the target heading (single) or the target heading accompanied by its surrounding neighbourhood of headings (multiple). The multiple setting enables the model's self-attention mechanism to leverage the structural context of the document better. While increasing the MAX\_LENGTH parameter to accommodate this context increases computational cost, it yields significant gains: we observed a substantial \textbf{10 percentage point improvement} in F1 score for data practice classification (e.g., distinguishing first party collection from third party sharing). However, the improvement for developer intent classification was marginal, showing only a 2 percentage point increase indicating the isolated headings still contain sufficient information to classify based on data practice intention.


\subsection{PrivSTRUCT versus PoliGraph}

\begin{figure}[t]
    \centering
    \includegraphics[width=0.5\textwidth]{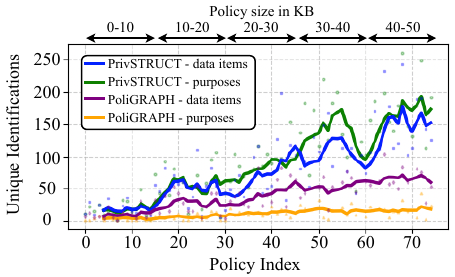}
    \caption{Comparison with PoliGraph}
    \label{fig:poligraph_compare}
\end{figure}

To benchmark our framework, we employ the open-source PoliGrapher NLP tool \cite{cui2023poligraph} to parse the 75 privacy policies in our test dataset ($DS_{test}$), comparing its extraction of data items and purposes against PrivSTRUCT. Figure \ref{fig:poligraph_compare} illustrates the comparative yield of unique item and purpose identifications.

Notably, the PoliGrapher tool failed to parse 6 of the 75 policies (8\%, all exceeding 25KB in size). While PrivSTRUCT successfully processed the entire dataset, we restrict the following comparative analysis strictly to the 69 policies successfully processed by both frameworks to ensure an equitable baseline and meaningful improvement metrics. In terms of data items, PrivSTRUCT identifies an average of 79.6 unique mentions per policy (accounting for multiple items embedded within single sentence excerpts), whereas PoliGrapher averages only 38.1 (a 52.1\% reduction). The disparity is even more pronounced for data purposes: PrivSTRUCT extracts an average of 92.8 unique purpose statements per policy, compared to PoliGrapher's 10.1 (an 89.1\% reduction). Furthermore, upon mapping these extracted excerpts to standardised purpose categories, PrivSTRUCT identifies an average of 2.9 additional distinct purpose categories per policy, demonstrating a significantly richer information yield.


\section{Results}
\label{5sec:Results}

For discussing the results of the large scale analysis of 3,756 privacy policy documents, we follow the metrics defined in Sub-sections ~\ref{subsec:metrics-compliance} and \ref{subsec:metrics-dilution}. We critically evaluate purpose compliance landscape based on each of the privacy policy (PP) and the data safety (DS) declaration of the most popular app representing the privacy policy. Given that both forms of disclosures (PP and DS) are developer-self declared, our expectation as well as with any end-user is to observe an explanatory text in the PP regarding the data purposes declared in DS. This is also enforced by regulatory frameworks and app market operator mandates under \emph{required consistency}. We observed that, despite the availability of 22 data item categories, majority of them were very sparse; for example, we did not or rarely observed any mention of \emph{d[5]-race/ethnicity} and \emph{d[6]-political/religious} categories. For the rest of this section, we discuss the purpose compliance with respect to eight high-frequent data item categories.

\begin{figure}[t]
    \centering
    \includegraphics[width=0.99\linewidth]{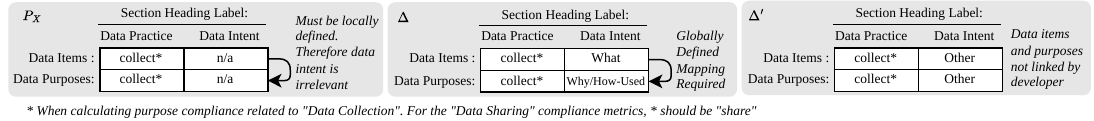}
    \caption{Locally-defined, globally-defined or un-defined/floating purposes}
    \label{5fig:supplementary}
\end{figure}

Figure \ref{5fig:supplementary} summarises how developers define data purposes compared to the data items in their disclosures. $P_X$ represents locally-defined purpose statements where it is irrelevant to specifically consider developer data intentions. They could be related to either first party collection or third party sharing based on the data practice classification label. $\Delta$ represents globally-defined purposes where an intention based mapping is required to link `what' data items are disclosed and `why' these data items are used. $\Delta'$ represents floating data purposes that are observed elsewhere in data intention classifications, yet still belonging to the same data practice category. 

\begin{figure}[t]
    \centering
    \includegraphics[width=0.99\textwidth]{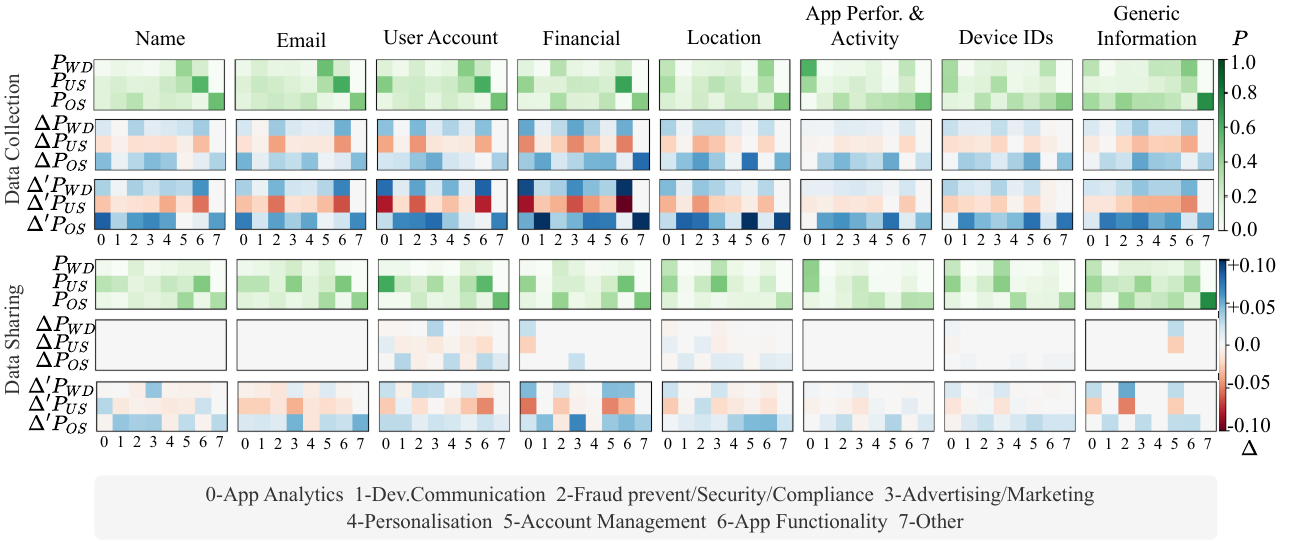}
    \caption{Results for Purpose Compliance}
    \label{fig:results_compliance_matrices}
\end{figure}

\begin{table}[t]
\centering
\caption{Average probability of occurrence for purpose well-disclosure(WD), under-statement(US) and over-statement(OS) for eight frequently used data item categories.}
\scriptsize
\begin{tabular}{llccccccccccc}
\hline
& &  &            & User &        &     & App   & Device & Gen- & \textbf{Avg.} &  & \\
& & Name & Email & Acct. & Finc. & Loc. & Acct. & IDs   & -eric & \textbf{$P_X$} &$\Delta$ &$\Delta'$\\
\hline
        &$P_{WD}$ & 0.111 & 0.154 & 0.159 & 0.122 & 0.143 & 0.144 & 0.135 & 0.187 & \textbf{0.144} & +0.017 & +0.026 ~(+18.0\%)\\
collect &$P_{US}$ & 0.214 & 0.230 & 0.278 & 0.253 & 0.196 & 0.143 & 0.165 & 0.137 & \textbf{0.202} & -0.016 & -0.026~(-12.9\%)\\
        &$P_{OS}$ & 0.222 & 0.204 & 0.190 & 0.199 & 0.214 & 0.264 & 0.244 & 0.347 & \textbf{0.235} & +0.028 & +0.048~(+20.4\%)\\
\hline
      & $P_{WD}$ & 0.073 & 0.099 & 0.055 & 0.063 & 0.097 & 0.080 & 0.099 & 0.131 & \textbf{0.087} & +0.001 & +0.006~(+6.9\%)\\
share & $P_{US}$ & 0.219 & 0.250 & 0.301 & 0.199 & 0.198 & 0.123 & 0.172 & 0.253 & \textbf{0.214} & -0.001 & -0.009~(-4.2\%)\\
      & $P_{OS}$ & 0.172 & 0.188 & 0.172 & 0.202 & 0.125 & 0.188 & 0.154 & 0.288 & \textbf{0.186} & +0.003 & +0.018~(+9.7\%)\\
\hline
\end{tabular}
\label{tab:purpose_compliance}
\end{table}

 \subsection{Well-and not so well-disclosed  Purposes}
 \label{subsec:results-purpose-compliance}

As shown in Figure \ref{fig:results_compliance_matrices}, we can observe that having a well-disclosed purpose for a given data item is not uniform (i.e., due to hot spots if we go through column-wise in the  row $P_{WD}$ for each data item). \emph{Name}, \emph{email}, and \emph{user account} have a higher probability of being well-disclosed when collected for account management. Similarly, \emph{financial} information is likely well-disclosed for fraud prevention/security and compliance and \emph{location, app performance, and app activity, device identifiers} for app analytics. All data items have a higher probability of being well-disclosed for app functionality. When \emph{location} and \emph{device identifier} are shared with third parties for advertising, it is likely to be well-disclosed.  

Table \ref{tab:purpose_compliance} quantitatively discuss the results based on each data when averaged across all data purpose categories. The probability of observing \textbf{well-disclosed and locally-defined} purposes among the selected data items remains low at 14.4\% for when collected and only 8.7\% when shared. Comparatively the purpose over-statement within the policy text remains quite high, approximately with twice the probability. Note that the three probability values in each column do not add up to 1 as we do not show purpose non-existence (which simply dictates that a data item and data purpose are unlikely to be linked via developer disclosures) here. 

Both in the figure and in table, we show the change to the original probabilities as soon as we include \textbf{globally-defined} purposes as $\Delta$. In the figure, blue shades represent the increase of the respective probabilities with red shades representing decreases. In the results, we also depict \textbf{unlinked/floating data purpose} compliance which are easy to be missed by end-users as the section headings do not properly disclose the developer intention with such data practices. We denote these contributions via $\Delta'$. As we identify more and more data-item linkages that are globally mapped, well-disclosures and over-statements tend to go higher while reducing the under-statements. More specifically, average $P_{WD}|collect$ improves by 18.0\% and average $P_{WD}|share$ does not increase as much; only by 6.9\%. This indicates that developers are less likely to emphasise on data they share with third parties, and rather tend to explain their collection usages. This is more apparent with nearly 40\% of a privacy policy is used for describing collection details and only 25\% for sharing details in average. Qualitatively we observed that policies tend to disclose redirections (e.g. `Refer to the third-party website's privacy policy for ...') instead of clearly indicating purposes. This raises a critical question for end-user privacy whether is it realistic for someone to read all such policies when many apps use third party services for analytics and advertising. The highest probability of understatement is observed with user account data being shared with third-parties.


\subsection{Purpose Dilution}
\begin{figure}[t]
    \centering
    \includegraphics[width=0.99\textwidth]{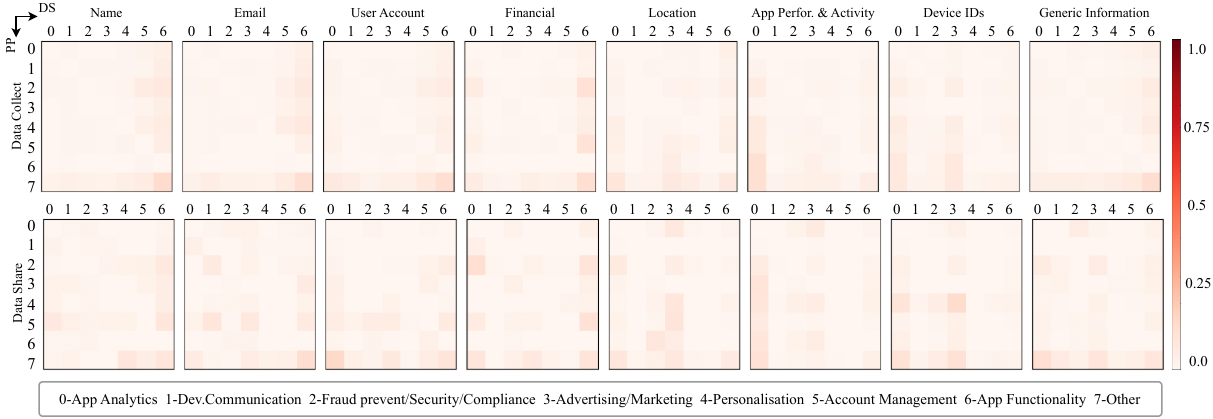}
    \caption{Results for Purpose Dilution}
    \vspace{-3mm}
    \label{fig:results_dilution_matrices}
\end{figure}

 Figure \ref{fig:results_dilution_matrices} represents the asymmetric purpose dilution matrices for the selected data items. As the data safety does not contain \emph{other} category, it is omitted from the x-axis of all matrices. However, ambiguous PP purposes could still be diluted across DS declarations. A significant purpose dilution can be characterised by horizontal or vertical lines that stand out in these matrices and hot spots are created when they intersect. 

 A primary observation is that \emph{Financial, location, app performance/activity, device identifier} data collected or shared for \emph{app analytics} is often diluted across other privacy policy purposes. A qualitative explanation is majority of privacy policies are not optimised based on app based services, rather they are optimised with online-services and traditional privacy policy architectures in mind. We argue that privacy policies representing mobile apps should require modifications or sections specifically mentioning mobile app based data practices. We also observe that \emph{device identifier} data collected or shared for \emph{advertising} is diluted as \emph{personalisation}, \emph{fraud prevention, security or compliance}, or \emph{other} purposes in the privacy policy. User \emph{email} when shared for \emph{advertising and developer communication} is considerably diluted as \emph{account management} which is a serious privacy concern for end-users. \emph{Financial details} when shared for \emph{analytics} are more easier to be misinterpreted as shared for \emph{fraud prevention and security and compliance purposes.} 
 Our tailored compliance metrics allow this finer grained evaluation and simply observing the hot spots in the figure, we could deduce that purpose dilution is more prominent among the data items that are shared with third parties. This again reinforces the discussion in Sub Section \ref{subsec:results-purpose-compliance} that more clearer explanations are required about third party data sharing within the privacy policies themselves.
 

 \section{Conclusion}

We presented \textbf{PrivSTRUCT}, a structure-aware framework designed to uncover the complex dependencies between data items and purposes in mobile app privacy policies. By moving beyond traditional flat-text analysis, we demonstrated that leveraging the hierarchical structure; specifically through the extraction and classification of section headings is essential for accurate interpretation, resolving a major limitation in prior work. Our methodological contributions include a hybrid architecture that pairs efficient encoder-based classifiers with a decoder-based information extractors. We showed that this approach not only enables cost-effective, locally hosted inference but also significantly outperforms existing state-of-the-art tools like PoliGraph, in average identifying  at least twice (or greater) as many data item or purpose excerpts in our benchmark comparisons.

 By leveraging robust extraction of data practices, items, and purposes from privacy policies, we characterise the disclosure landscape of popular apps in the Google Play Store. We uncover the relationships between globally defined purposes and specific data items. 
 Our analysis reveals that, compared to relying solely on locally defined purposes, the probability of developers overstating purposes is 20.4\% higher for first-party collection and 9.7\% higher for third-party sharing.
 We attribute this to developers relying on `global' purpose definitions that are difficult for end-users to parse at a glance. Alarmingly, this encourages blanket consenting, as the probability of adequate purpose disclosure remains critically low.


Most critically, our analysis of ``Purpose Dilution'' highlights that third-party data sharing remains the most opaque aspect of the ecosystem. We observed that highly sensitive data flows, such as the sharing of financial information or device identifiers for advertising, are frequently diluted into unrelated categories like ``Fraud Prevention'' or generic ``Other'' clauses within the policy text. This misalignment suggests that despite the introduction of simplified Data Safety labels, the underlying legal documents remain a barrier to true transparency.

\begin{credits}
\subsubsection{\ackname} This research was funded by the Australian Research Council (ARC) Discovery Project (DP220102520).

\subsubsection{\discintname}
The authors have no competing interests to declare that are relevant to the content of this article.
\end{credits}
%
%
%
\bibliographystyle{splncs04}
\bibliography{reference}
%





\appendix
\section{Appendix}

\subsection{DPO for heading extraction}
\label{appendix:DPO}

\begin{figure}[H]
    \centering
    \includegraphics[width=0.6\textwidth]{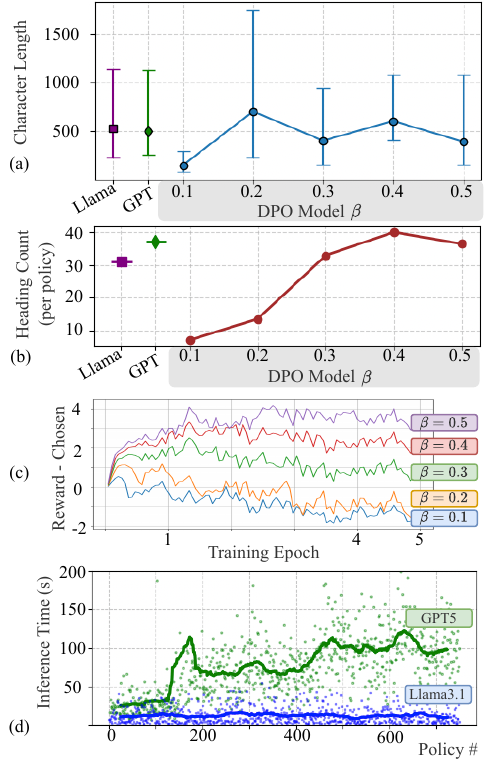}
    \caption{DPO training results for heading extraction.}
    \label{fig:dpo_results}
\end{figure}

We show the benchmarking results of the Direct Preference Optimisation (DPO) training in Figure \ref{fig:dpo_results}. Sub Figure (a) represents the fitment of the identified headings across the policy based on IQR and median of the accompanied content of each heading for five $\beta$ values. We would like to refer the reader to sub figure (b) at the same time, where we depict the average number of headings we identify for each $\beta$. We observed that $\beta=0.1$ the default parameter for the \texttt{DPO Trainer}—proved too aggressive for our task, resulting in a collapse where the model failed to identify meaningful headings. Comparing our DPO-Llama3.1 model against the baseline references (standard Llama3.1 Instruct and GPT-5), we observe that GPT-5 generally yields tighter IQRs and identifies a higher volume of headings. However, as we increase $\beta$ to encourage the model not to deviate excessively from the reference policy, stability improves. Specifically, $\beta=0.4$ offers the optimal balance, achieving an alignment median closest to both Llama3.1 and GPT-5, while maintaining a lower quartile distinct from zero (a zero value typically indicates list items with no subsequent content misclassified as headings). Sub figure (c) further emphasises the instability at lower values ($\beta=0.1, 0.2$), highlighting the necessity of hyper-parameter tuning over relying on defaults for structural extraction tasks.

Additionally, we conducted a manual evaluation for the headings we identified, indicating the author preference for the outputs when comparing respective DPO-trained Llama3.1, standard Llama3.1 and GPT5 models. Out of the 75 policies we compare among the test dataset, we preferred GPT outputs 50.67\% followed up by DPO-Llama at 42.67\%. This result is significant as the latter performed well given it a much smaller decoder model compared to the state-of-the-art GPT5 with reasoning capabilities. Moreover, simply compared with Llama3.1 we preferred the DPO-Llama3.1, 65.33\% of the times. This benchmarking depicts that, \emph{we can reliably use a DPO trained locally hosted decoder models for heading extraction with no API costs and much faster inference times (c.f. sub figure d)}. For the at-scale analysis we use the DPO trained Llama model for heading-extraction task.  

\end{document}